\documentclass[conference,letterpaper]{IEEEtran}

\usepackage[utf8]{inputenc} 
\usepackage[T1]{fontenc}
\usepackage{url}
\usepackage{ifthen}
\usepackage{cite}
\usepackage[cmex10]{amsmath} % Use the [cmex10] option to ensure complicance
\usepackage{textcomp,enumerate, url, cancel}                    			% Document and Style 
\usepackage[margin=0.8in]{geometry}
\usepackage{color}
\usepackage{amssymb,amsthm,bbm,bm,nicefrac}   				% Mathematical symbols
\usepackage{graphicx,epsfig}                                                               		% Graphics
\usepackage{multirow}
\newcommand{\ket}[1]{\left|#1\right\rangle}

\newcommand{\braket}[2]{\left\langle #1\lvert#2\right\rangle}
\newcommand{\abs}[1]{\left\lvert #1\right\rvert}

\newcommand{\dketbra}[1]{\left|#1\right\rangle\left\langle#1\right|}

\newcommand{\be}{\begin{equation}} 							
\newcommand{\ee}{\end{equation}}
\newcommand{\ba}{\begin{align}}
\newcommand{\ea}{\end{align}}
\newcommand{\bematrix}{\left(\begin{matrix}}
\newcommand{\ematrix}{\end{matrix}\right)}

\theoremstyle{definition}

\theoremstyle{theorem}
\newtheorem{theorem}{Theorem}[section]

\theoremstyle{lemma}

\theoremstyle{proposition}

\theoremstyle{corollary}

\theoremstyle{observation}

\theoremstyle{remark}

\def\ii{\mathrm{i}}
\newcommand{\tr}[1]{\mathrm{tr}\left[#1\right]}

\def\cA{\mathcal A}

\def\cN{\mathcal N}

\def\Do{\Delta\omega}
\def\Dt{\Delta\theta}
\def\sinc{{\rm sinc}}

\begin{document}
\title{Performance of Coherent Frequency-Shift Keying for Classical Communication \\ on Quantum Channels } 

% %%% Single author, or several authors with same affiliation:
% \author{%
%   \IEEEauthorblockN{Stefan M.~Moser}
%   \IEEEauthorblockA{ETH Zürich\\
%                     ISI (D-ITET)\\
%                     CH-8092 Zürich, Switzerland\\
%                     Email: moser@isi.ee.ethz.ch}
% }

%%% Several authors with up to three affiliations:
\author{%
  \IEEEauthorblockN{Matteo Rosati}
\IEEEauthorblockA{\textit{Departament de F\'{\i}sica: Grup d'Informaci\'{o} Qu\`{a}ntica} \\
\textit{Universitat Aut\`onoma de Barcelona}\\
Bellaterra (Barcelona),
Spain \\
matteo.rosati@uab.cat}
}

%%% Many authors with many affiliations:
% \author{%
%   \IEEEauthorblockN{Albus Dumbledore\IEEEauthorrefmark{1},
%                     Olympe Maxime\IEEEauthorrefmark{2},
%                     Stefan M.~Moser\IEEEauthorrefmark{3}\IEEEauthorrefmark{4},
%                     and Harry Potter\IEEEauthorrefmark{1}}
%   \IEEEauthorblockA{\IEEEauthorrefmark{1}%
%                     Hogwarts School of Witchcraft and Wizardry,
%                     1714 Hogsmeade, Scotland,
%                     \{dumbledore, potter\}@hogwarts.edu}
%   \IEEEauthorblockA{\IEEEauthorrefmark{2}%
%                     Beauxbatons Academy of Magic,
%                     1290 Pyrénées, France,
%                     maxime@beauxbatons.edu}
%   \IEEEauthorblockA{\IEEEauthorrefmark{3}%
%                     ETH Zürich, ISI (D-ITET), ETH Zentrum, 
%                     CH-8092 Zürich, Switzerland,
%                     moser@isi.ee.ethz.ch}
%   \IEEEauthorblockA{\IEEEauthorrefmark{4}%
%                     National Chiao Tung University (NCTU), 
%                     Hsinchu, Taiwan,
%                     moser@isi.ee.ethz.ch}
% }

\maketitle

%%%%%%
%% Abstract: 
%% If your paper is eligible for the student paper award, please add
%% the comment "THIS PAPER IS ELIGIBLE FOR THE STUDENT PAPER
%% AWARD." as a first line in the abstract. 
%% For the final version of the accepted paper, please do not forget
%% to remove this comment!
%%
\begin{abstract}
We evaluate the performance of coherent frequency-shift keying (CFSK)~\cite{Burenkov2018a,Burenkov2020} alphabets for communication on quantum channels. We show that, contrarily to what previously thought, the square-root-measurement (SRM) is sub-optimal for discriminating CFSK states.
Furthermore, we compute the maximum information transmission rate of the CFSK alphabet, observing that it employs at least as many frequency modes as the signal states, and compare it with standard phase-shift-keying. 
Finally, we introduce a discretized CFSK alphabet with higher mode-efficiency, exhibiting comparable error-probability performance with respect to CFSK and better rate performance. Our results suggest the existence of a tradeoff between the CFSK reduced error-probability and its mode efficiency. 
\end{abstract}

%% The paper must be self-contained. However, if you are referring to
%% a full version for checking certain proofs, please provide the
%% publically accessible location below.  If the paper is completely
%% self-contained, you can remove the following line from your
%% submission.

\section{Introduction}
The recently introduced coherent frequency-shift keying (CFSK) alphabet has shown promising potential to decrease the error rate in long-distance communication on quantum channels~\cite{Burenkov2018a,Burenkov2020}. In particular, in Ref.~\cite{Burenkov2020} it was shown experimentally that the error probability attainable with CFSK and adaptive detection can outperform the standard phase-shift keying (PSK) alphabet. Here we provide an initial analysis of CFSK for communication purposes, taking into account both the optimal discrimination error probability and the achievable information-transmission rate. 

\section{Coherent frequency-shift keying alphabet}
The CFSK alphabet, ${\cal A}_{\rm cfsk}:=\{\ket{\alpha_{m}}\}_{m=0}^{M-1}$, comprises coherent-state signals with variable shape in the time and frequency domain, of the form
\be
\ket{\alpha}:=e^{-\frac{\abs{\alpha}^{2}}{2}}\sum_{n=0}^{\infty}\frac{1}{n!}\left[\int d\omega \tilde\alpha(\omega)a^{\dag}(\omega)\right]^{\otimes n}\ket0,
\ee
where $a^{\dag}(\omega)$ is the photon-creation operator at frequency $\omega$, $\ket0$ is a tensor-product of vacuum states at all frequencies and $|\alpha|^{2}$ is the total number of photons of the signal. The specific frequency-shape can be determined by Fourier-transform of the time-shape: $\tilde\alpha(\omega):=\int \frac{dt}{\sqrt{2\pi}} \alpha(t) e^{\ii \omega t}$. In particular, CFSK signals have a temporal shape 
\be
\alpha_{m}(t):=\alpha f(t)e^{-\ii (\theta_{m}+\omega_{m}t)},
\ee
with $\theta_{m}=m\Dt$ and $\omega_{m}=\omega_{0}+ m\Do$. Note that the total-photon-number constraint $|\alpha|^{2}=\int dt |\alpha_{m}(t)|^{2}$ implies the normalization $\int dt |f(t)|^{2}=1$. 

A key object in evaluating the discrimination error and Holevo rate of an alphabet is the Gram matrix $G$, whose elements are given by the states' overlaps. For the CFSK alphabet we have
\be
G_{jk}:=\braket{\alpha_{j}}{\alpha_{k}}=e^{-|\alpha|^{2}}\sum_{n=0}^{\infty}\frac{\left[\int d\omega \tilde\alpha_{j}(\omega)^{*}\tilde\alpha_{k}(\omega)\right]^{n}
}{n!},
\ee
where we used that $[a(\omega),a^{\dag}(\omega')]=\delta(\omega-\omega')$. The integral can be computed as follows:
\begin{align}
\int d\omega \tilde\alpha_{j}(\omega)^{*}&\tilde\alpha_{k}(\omega)=|\alpha|^{2}\int d\omega \frac{dt dt'}{2\pi} f(t)f(t') \cdot \\
&\cdot e^{-\ii [(k-j)\Dt + (\omega_{k}-\omega) t - (\omega_{j}-\omega)  t')]}\\
&=|\alpha|^{2}\int dt |f(t)|^{2} e^{-\ii (k-j)(\Dt+\Do t)}\\
&=|\alpha|^{2}e^{\ii (j-k)\Dt}\sqrt{2\pi}\tilde F((j-k)\Do),
\end{align}
having defined $F(t):=|f(t)|^{2}$. Hence the CFSK Gram matrix is a positive-definite Toeplitz matrix, i.e., whose elements depend only on the difference between their indices:
\be\small\label{eq:gramEl}
G_{jk}=g_{j-k}=\exp[-\abs{\alpha}^{2}\left(1-e^{\ii (j-k)\Dt}\sqrt{2\pi}\tilde F((j-k)\Do)\right)].
\ee
In the original proposal~\cite{Burenkov2018a,Burenkov2020}, the authors consider a rectangular shape of duration $T$: $f_{\rm r}(t):=\frac{1}{\sqrt T}$ for $t\in[0,T]$ and zero otherwise, so that 
\begin{align}
&\tilde\alpha_{m}(\omega)=\alpha\sqrt{\frac{T}{2\pi}}e^{-\ii\theta_{m}+\ii \frac{\omega-\omega_{m}}{2}T}{\rm sinc}\left( \frac{\omega-\omega_{m}}{2}T\right),\\
&\tilde F_{\rm r}(\omega)=\frac{e^{\ii \frac{\omega T}{2}}}{\sqrt{2\pi}}\sinc\left(\frac{\omega T}{2}\right).
\end{align} Moreover, the values of the parameters are restricted to the range $\Dt,\Do T\in[0,2\pi)$. Note that for $\Dt=2\pi/M$ and $\Do=0$ we obtain the standard phase-shift keying (PSK) alphabet with circulant Gram matrix.

\section{Sub-optimality of the square-root-measurement}
The square-root-measurement (SRM) is a well-known quantum measurement employed in communication and state discrimination~\cite{Hausladen1994,Hausladen1996}. It can be constructed from the Gram matrix of the states as follows~\cite{Ban1997}: let $\Phi:=\sum_{m=0}^{M-1}\dketbra{\alpha_{m}}$ be the sum of all states in the alphabet $\cA$; then the SRM is comprised of rank-one projectors $\ket{\mu_{m}}:=\Phi^{-1/2}\ket{\alpha_{m}}$. The Gram matrix is exactly the matrix representation of $\Phi$ in the basis $\{\ket{\mu_{m}}\}_{m=1}^{M}$. Then the SRM success probability in discriminating the states is
\be\small
P_{\rm succ}^{\rm srm}(\cA)=\frac1M\sum_{m=0}^{M-1}|\braket{\mu_{m}}{\alpha_{m}}|^{2}=\frac1M\sum_{m=0}^{M-1}\abs{(G^{1/2})_{mm}}^{2}.
\ee

The conditions for its optimality in state discrimination have been recently formulated in terms of the Gram matrix of the states to be discriminated:
\begin{theorem}[Optimality of the SRM~\cite{Pozza2015}]\label{thm:srm}
The SRM is optimal for the minimum-error discrimination of a set of states with Gram matrix $G$ if and only $(G^{1/2})_{mm}$ is constant for all $m$.
\end{theorem}

\begin{figure}[t!]
\center\includegraphics[scale=.63]{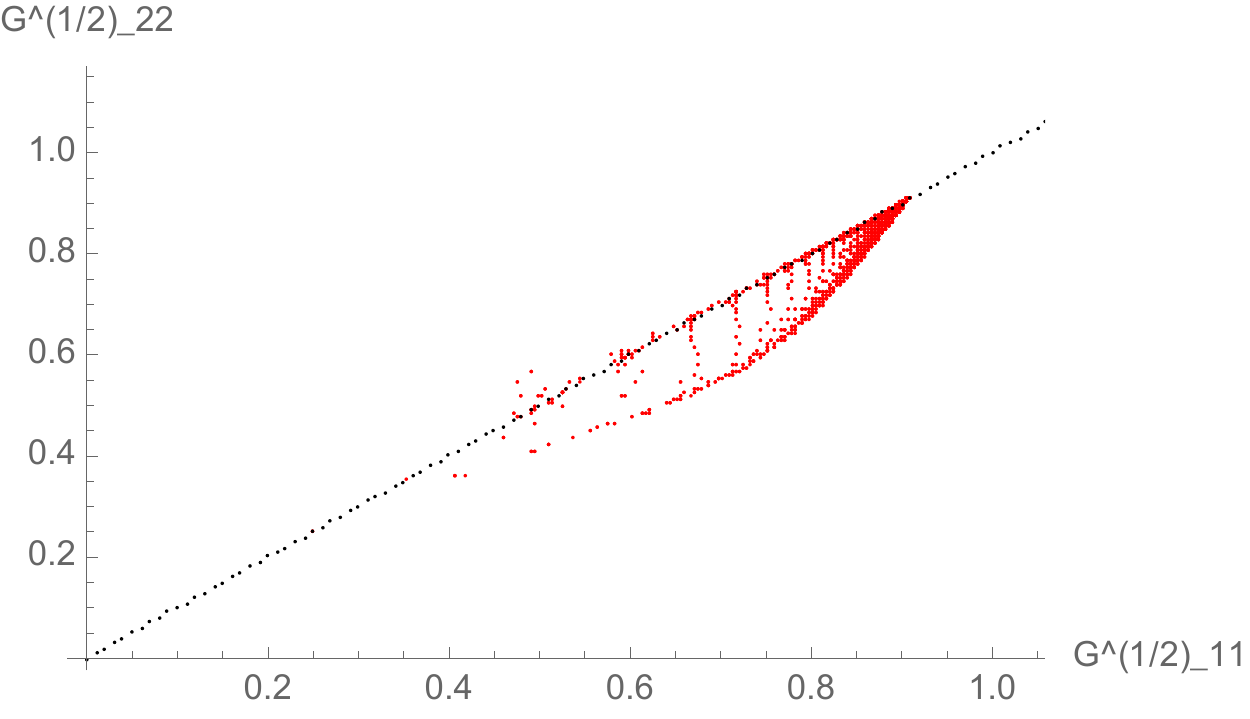}\\ \center\includegraphics[scale=.3]{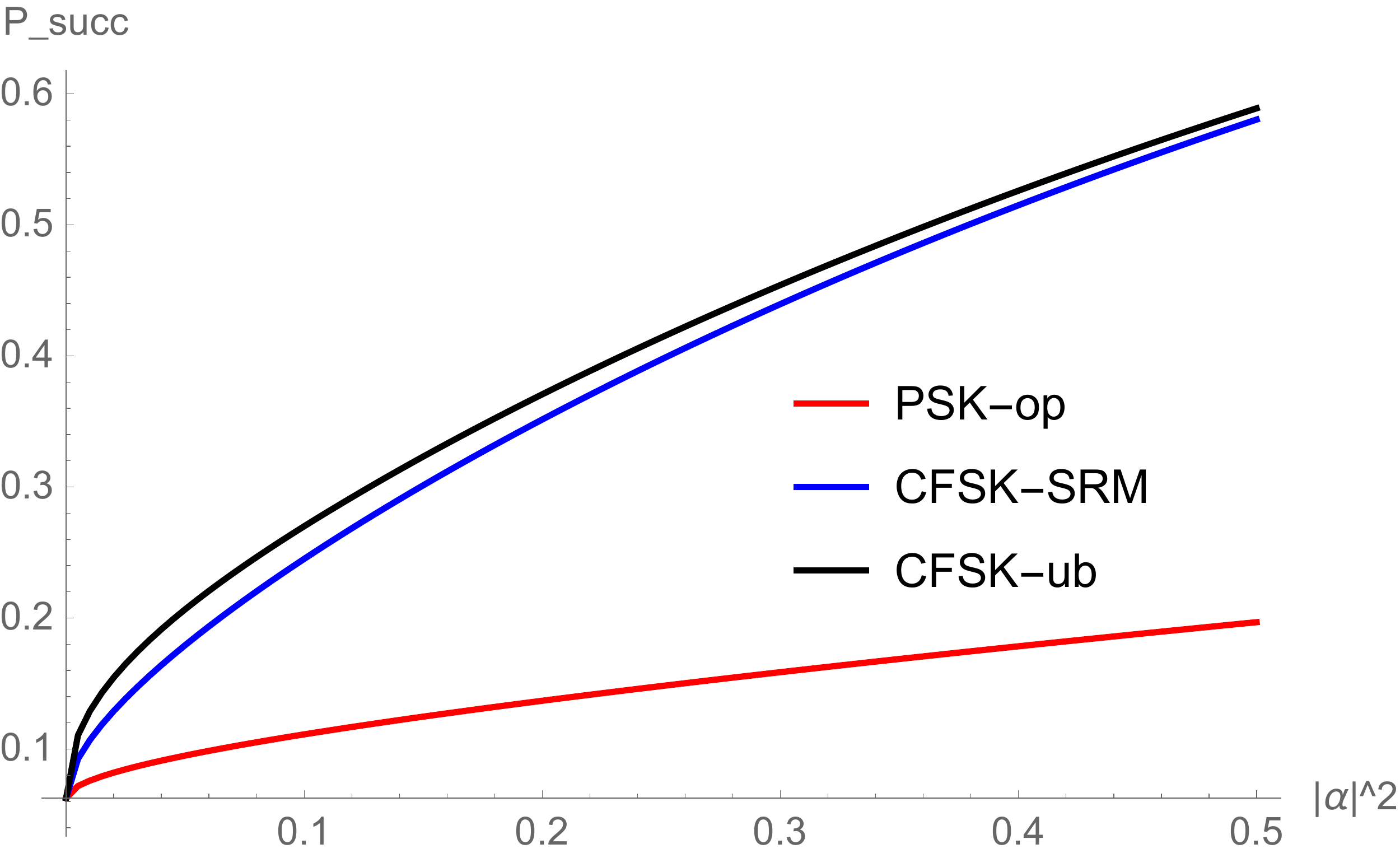}
\caption{Top - Plot of points in the plane $((G^{1/2})_{11},(G^{1/2})_{22})$, as determined by the CFSK alphabet for $M=16$ with several values of $\Dt,\Do T\in[0,2\pi)$ (red), and of the line $G^{1/2}_{11}=G^{1/2}_{22}$ (black). Bottom - Plot of the upper bound~\cite{Sentis2016} (black) on the optimal CFSK success probability, the SRM (blue) success probability for CFSK and the optimal success probability for PSK (red). The alphabet size is $M=16$ and the values of $\Dt, \Do T$ for CFSK are the optimal ones of Ref.~\cite{Burenkov2020}.}
\label{fig1}
\end{figure}

It is easy to check numerically that the Gram matrix of the CFSK states in general does not satisfy the theorem above, hence the SRM cannot be optimal for these states, contrarily to what stated in Refs.~\cite{Burenkov2018a,Burenkov2020}, see Fig.~\ref{fig1}(top). Nevertheless, Ref.~\cite{Sentis2016} provides general upper and lower bounds on the optimal error probability:
\begin{align}
&P_{\rm succ}^{\rm op}(\cA)\leq\left(\frac{\tr{G^{1/2}}}{M}\right)^{2}+\sqrt{\gamma_{\rm max}}\|{\bf q}-{\bf u}\|_{1},\\
&P_{\rm succ}^{\rm op}(\cA)\geq P_{\rm succ}^{\rm srm}(\cA)\geq\left(\frac{\tr{G^{1/2}}}{M}\right)^{2},
\end{align}
where $q_{m}:=\frac{(G^{1/2})_{mm}}{\tr{G^{1/2}}}$, $u_{m}:=\frac1M$ and $\gamma_{\rm max}$ is the maximum eigenvalue of $G$. A numerical calculation of the upper bound shows that the SRM might be underperforming with respect to the optimal measurement, especially at low energies, see Fig.~\ref{fig1}(bottom).
%other numerics...

\begin{figure}[t!]
\center\includegraphics[scale=.3]{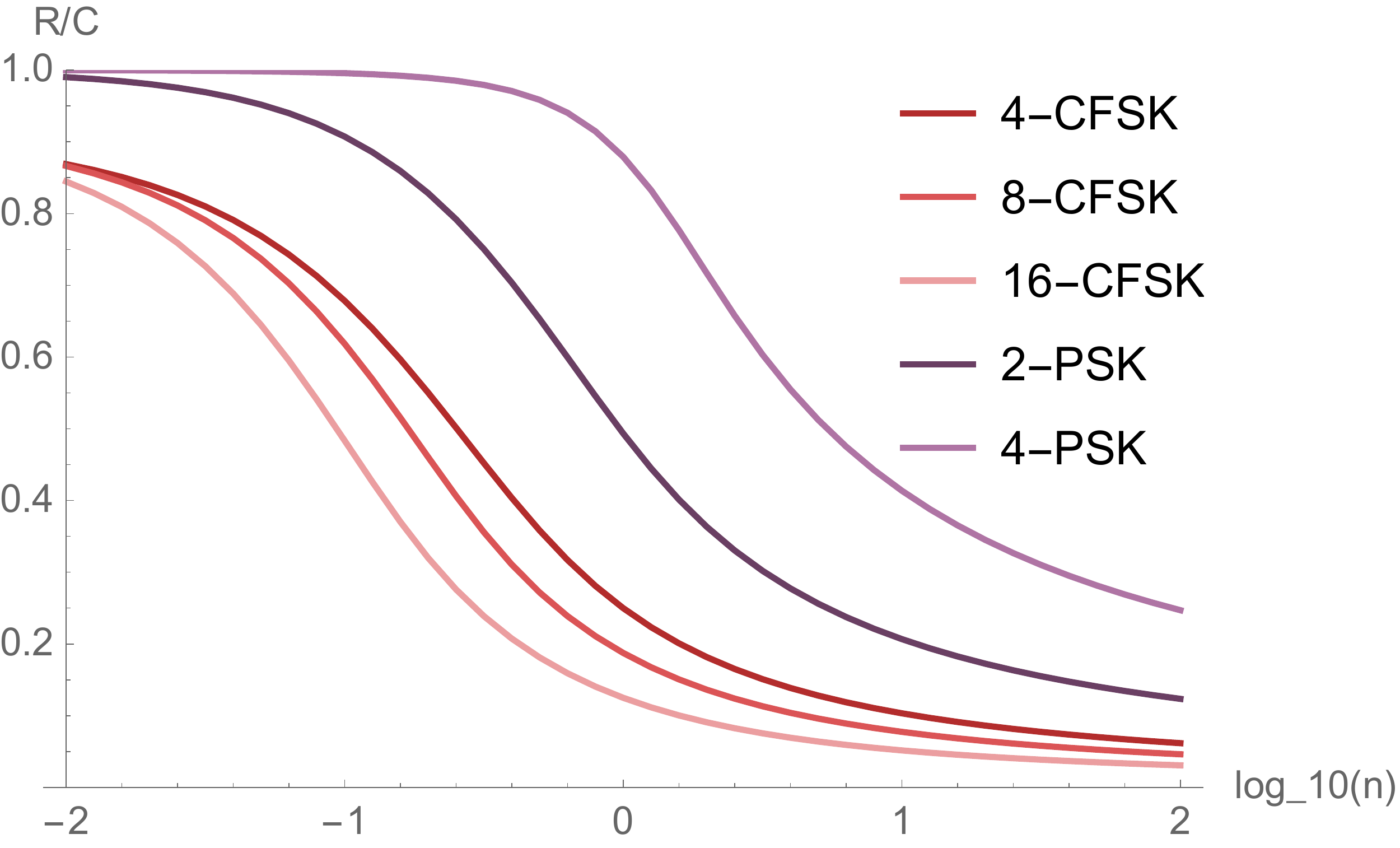}
\caption{Plot of the ratio $R/C(n)$ between CFSK, PSK maximum rates and the channel capacity at the corresponding average energy, for several values of alphabet size. For CFSK, we set the optimal values of $\Dt$, $\Do T$ for each $M$ provided in Ref.~\cite{Burenkov2020}.}
\label{fig2}
\end{figure}

\section{Communication rate}
Note that knowing the Gram matrix, in particular its spectrum, also allows a straightforward calculation of the Holevo rate of the alphabet, i.e., the maximum information-transmission rate attainable asymptotically by sending typical sequences of signals picked from $\cA_{\rm cfsk}$ and decoding them with collective quantum measurements~\cite{HolevoBOOK,Rosati16b}. Indeed we have
\be\begin{aligned}\label{eq:hol}
\chi(\cA)&:=H\left(\frac{\Phi}{M}\right)-\sum_{m=0}^{M-1}H(\dketbra{\alpha_{m}})\\
&=H\left(\frac{G}{M}\right),
\end{aligned}\ee
where $H(\cdot)$ is the Von Neumann entropy in base $2$. 
%We compare this quantity with the broadband capacity of the channel~\cite{Giovannetti2003}:
%\be
%C_{\rm bb}(E)=\frac{1}{\ln 2}\sqrt{\frac{\pi E}{3\hbar}},
%\ee

This quantity determines the maximum rate of the classical-quantum channel induced by the CFSK alphabet:
 \begin{equation}
 \cN_{\rm cfsk}: m\in\{0,\cdots,M-1\}\mapsto \ket{\alpha_{m}}\in\cA_{\rm cfsk}.
 \end{equation}
Indeed, since the corresponding signals are centered around $M$ distinct frequencies, we have to consider that $\cN_{\rm cfsk}$ uses at least $M$ parallel and distinct frequency modes (or uses of the communication line). Hence, fixing the average number of photons per mode employed by $\cN_{\rm cfsk}$ to $n$, the constraint on the total number of photons of each signal reads $|\alpha|^{2}=M n$. The mode efficiency, i.e., maximum number of transmitted bits per second per mode, of the CFSK channel at average photon number per mode $n$ then is 
 \begin{equation}
 R_{\rm cfsk}(n;M):=\frac1M\chi(\cA_{\rm cfsk})\Big|_{|\alpha|^{2}=M n}.
 \end{equation}
 
%where $E:=\frac1M\sum_{m=0}^{M-1}E_{m}$ is the average energy of the alphabet and $E_{m}$ the mean energy of a single letter:
%\be\begin{aligned}
%E_{m}&=\int d\omega \hbar\omega \abs{\tilde\alpha_{m}(\omega)}^{2}\\
%&=\abs{\alpha}^{2}\frac{T}{2\pi} \int d\omega\,\hbar\omega\,{\rm sinc}^{2}\left( \frac{\omega-\omega_{m}}{2}T\right)\\
%&=\abs{\alpha}^{2}\frac{T}{2\pi} \hbar\omega_{m} \int d\omega\,{\rm sinc}^{2}\left( \frac{\omega-\omega_{m}}{2}T\right)\\
%&=\abs{\alpha}^{2}\hbar\omega_{m}.
%\end{aligned}\ee
%The resulting average energy of the alphabet then is
%\be\label{eq:aveEn}
%E=\abs{\alpha}^{2}\hbar\left(\omega+\Do\frac{M-1}{2}\right).
%\ee
In Fig.~\ref{fig2} we compare this quantity with the narrowband channel capacity of a lossy bosonic channel~\cite{Giovannetti2004}, modeling optical fiber and free-space links, with a received number of photons per mode equal to $n$,
\be
C(n):=(n+1)\log_{2}(n+1)-n\log_{2} n.
\ee
Moreover, we compare with the maximum rate attainable by a PSK constellation~\cite{Kato1999}. We observe that the superior discrimination performance of the CFSK alphabet with respect to the PSK one does not translate in a superior communication performance. This might be due to the fact that PSK states can occupy a single frequency mode, i.e., a single use of the communication line, whereas CFSK states require at least $M$ distinct modes to transmit a single signal. Hence, CFSK states appear to make an inefficient use of the available bandwidth, which increases their distinguishability. In order to investigate this delicate issue further, in the next section we provide a narrowband approximation of the CFSK alphabet, which allows to tune the number of occupied modes independently of the number of signals.
%small alfa large M

\section{Discrete-mode alphabet}
Here we consider a simplified alphabet, named discrete CFSK (dCFSK), that aims at reproducing the main features of CFSK within a narrowband discrete-mode setting. The new alphabet comprises $M$ signals on $2L+1$ channel uses, $\cA_{\rm dcfsk}(L):=\{\ket{\vec\alpha_{m}}:=\otimes_{\ell=-L}^{L}\ket{\alpha_{m,\ell}}\}_{m=0}^{M-1}$, where each signal is a product of single-mode coherent states with amplitudes $\alpha_{m,\ell}:=a_{\ell}e^{\ii\theta_{m,\ell}}$, where $a_{\ell}\geq0$ and $\theta_{m,\ell}\in[0,2\pi)$, so that $\vec\alpha_{m}=(\alpha_{m,\ell})_{\ell=-L}^{L}$. The total number of photons is constant and equal to $a^{2}:=\sum_{\ell=-L}^{L}a_{\ell}^{2}$. The Gram matrix of the states is
\be\begin{aligned}
G'_{m,n}(a)&=\exp-\sum_{\ell=-L}^{L}a_{\ell}^{2}\left[1-e^{\ii(\theta_{m,\ell}-\theta_{n,\ell})}\right]\\
&=\exp-a^{2}\left[1-\sum_{\ell=-L}^{L}F_{\ell}e^{\ii(\theta_{m,\ell}-\theta_{n,\ell})}\right],
\end{aligned}\ee
so that, by setting $\theta_{m,\ell}=m\cdot(\Dt+\ell\cdot \Delta )$ and controlling the energy distribution on each mode through the fractions $F_{\ell}:=(a_{\ell}/a)^{2}$, with $\sum_{\ell=-L}^{L}F_{\ell}=1$, we can approximate the Gram matrix of CFSK states, Eq.~\eqref{eq:gramEl}. Indeed, let us define the $L$-th order Fourier series expansion of a function $\tilde F(t)$ as
\begin{align}
&S_{L}(t):=\sum_{\ell=-L}^{L}c_{\ell}e^{\ii t\ell \Delta},\\
&c_{\ell}:=\left(\frac{\Delta}{2\pi}\right)\int_{-\frac\pi \Delta}^{\frac\pi \Delta}dt \tilde F(t) e^{-\ii t \ell \Delta},
\end{align}
with $\Delta=\pi/(M-1)$. Then, for increasing $L$, the sums $S_{L}(t)$ provide increasingly better approximations of $F(t)$ in the interval $t\in[-M-1,M-1]$, provided that $F(t)$ satisfies a sufficient condition for the convergence of the Fourier series, e.g., differentiability at every $t$. 
Consequently, when the coefficients $c_{\ell}$ are positive we can normalize them by $S_{L}(0)=\sum_{\ell=-L}^{L}c_{\ell}$ and identify $F_{\ell}=c_{\ell}/S_{L}(0)$, obtaining
\be
\tilde{G}_{m,n}(a)=\exp-a^{2}\left[1-e^{\ii\Dt(m-n)}\frac{S_{L}(m-n)}{S_{L}(0)}\right],
\ee
which approximates the CFSK Gram matrix \eqref{eq:gramEl}. Using $\Delta\theta'=\Delta\theta+\pi/2$ and the shape $F(t)=\sinc(t \Delta\omega T/2)$ in order to reconstruct the shape described above, one can evaluate success probabilities and information transmission rates of $\cA'_{\rm cfsk}(L)$, for $L$ and $M$ such that the corresponding Fourier coefficients are positive. 

\begin{figure}[t!]
\center\includegraphics[scale=.3]{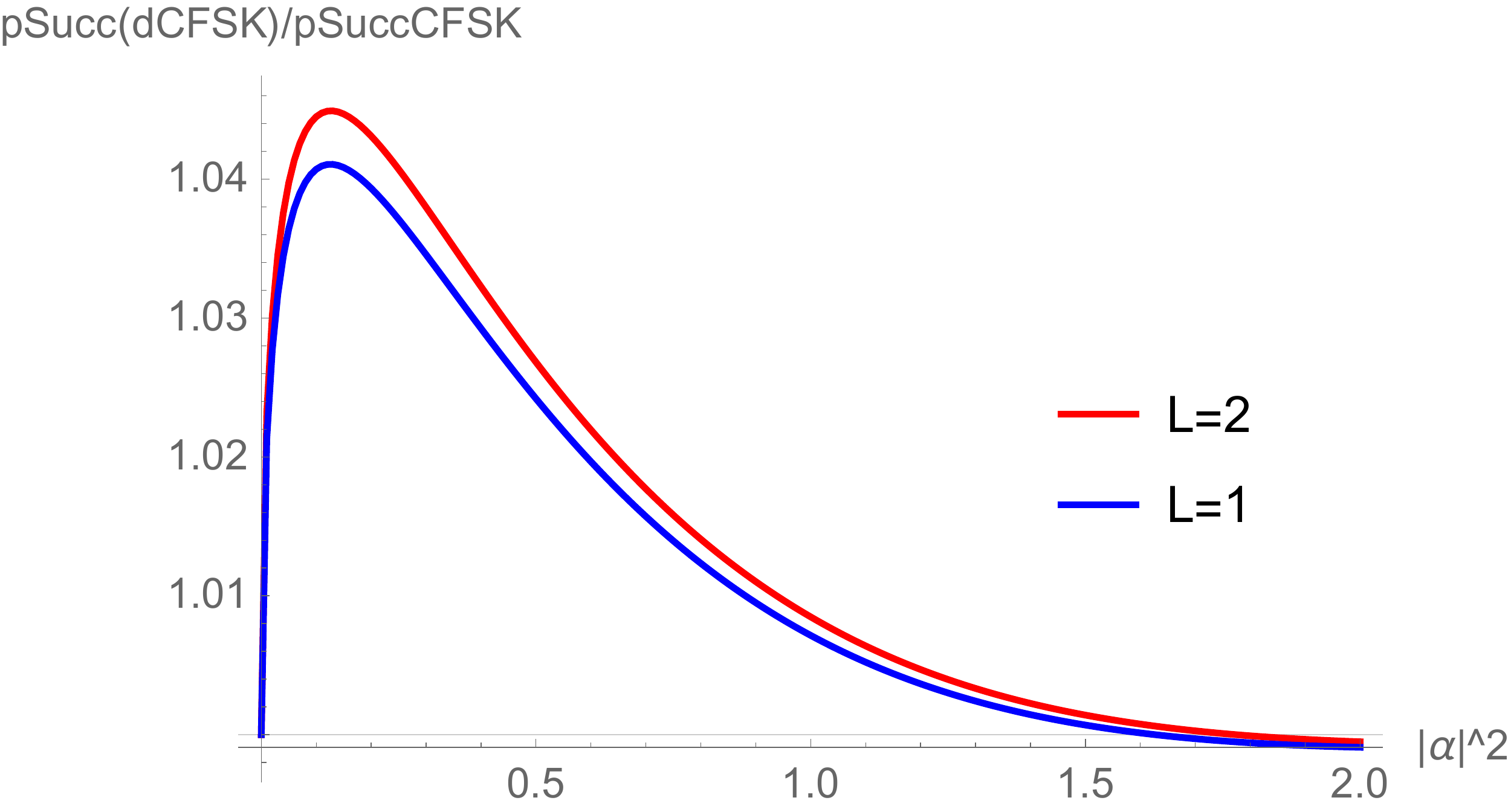}
\caption{Plot of the ratio between the SRM success probability of $\cA_{\rm dcfsk}$, for several values of $L$, and that of other alphabets $\cA_{\rm cfsk}$ for $M=4$. The parameter values are the same of Fig.~\ref{fig2}.}
\label{fig3}
\end{figure}

In Fig.~\ref{fig3} we show the ratio between the SRM success probability of the dCFSK alphabet and the CFSK one for $M=4$ and $L=1,2$. Already $L=1$ is sufficient to reproduce the same discrimination performance of CFSK with dCFSK, and even slightly surpass it. Observe that, as in the case of $\cA_{\rm cfsk}$, the SRM for $\cA_{\rm dcfsk}(L)$ will not be optimal in general and a further increase of the advantage over PSK can be expected. 

\begin{figure}[t!]
\center\includegraphics[scale=.3]{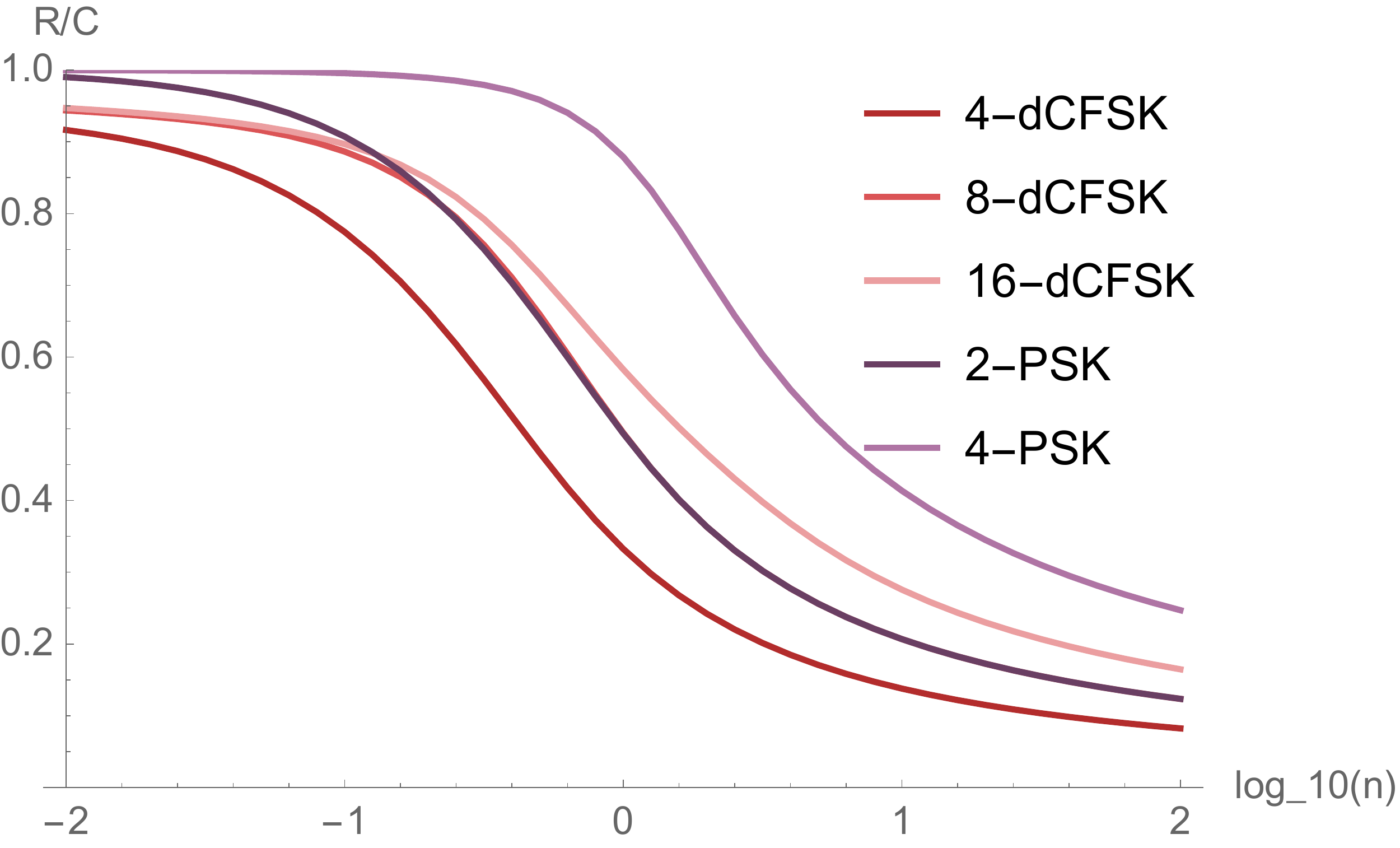}
\caption{Plot of the ratio $R/C(n)$ between dCFSK ($L=1$), PSK maximum rates and the channel capacity at the corresponding average energy, for several values of alphabet size.}
\label{fig4}
\end{figure}

The advantage of the dCFSK alphabet is that we can tune the number of modes it employs, independently of $M$. To study this potential, we then compute the Holevo rate of the dCFSK alphabet, similarly to Eq.~\eqref{eq:hol}, and define the dCFSK channel
 \begin{equation}
 \cN_{\rm dcfsk}: m\in\{0,\cdots,M-1\}\mapsto \ket{\vec\alpha_{m}}\in\cA_{\rm dcfsk}.
 \end{equation}
 Finally, we compute the mode-efficiency of this channel, which employs $2L+1$ parallel and distinct modes, with an average photon number per mode $n$ and total photon number per signal $a^{2}= (2L+1) n$:
 \begin{equation}
 R_{\rm dcfsk}(n;M;L):=\frac{1}{(2L+1)}\chi(\cA_{\rm dcfsk})\Big|_{|\alpha|^{2}=(2L+1) n}.
 \end{equation}
 A comparison of this rate for $L=1$ with PSK rates, per unit of channel capacity is shown in Fig.~\ref{fig4}. The performance of the dCFSK alphabet clearly improves with respect to the CFSK one, showing now an increase of the communication rate as $M$ increases, and it is even able to beat a binary PSK modulation. However, by increasing the size of the PSK alphabet we can still always surpass dCFSK rates.
 
%capacities discrete. 

% explicit detection scheme?

\section{Discussion and conclusions}
We have analyzed the error probability and maximum information transmission rate of the recently introduced CFSK alphabet. For this alphabet, we showed that the gain in success probability with respect to PSK can be enhanced beyond the SRM limit, contrarily to what previously thought in Refs.~\cite{Burenkov2018a,Burenkov2020}. Moreover, we observed that such gain comes at the cost of a reduced mode- (or bandwidth-) efficiency of the alphabet for communication purposes, which seriously decreases its communication rate with respect to PSK. 

We then introduced a discrete version of CFSK where the number of occupied modes can be tuned at will, which maintains the CFSK advantage in discrimination. At the same time, we showed that the dCFSK alphabet has enhanced communication performance with respect to CFSK. Still, in absolute terms, the PSK alphabet seems to be always advantageous for a sufficiently large number of signals. 

Future work will focus on further studying the potential of dCFSK constellations, searching for other possible adaptions of the alphabet with better communication performance and analyzing practical decoder designs for the discrete case.

\section*{Acknowledgments}
This project has received funding from the European Union's Horizon 2020 research and innovation programme under the Marie Sk\l odowska-Curie grant agreement No 845255. We acknowledge useful discussions with J. Calsamiglia and A. Winter.

%%%%%%
%% To balance the columns at the last page of the paper use this
%% command:
%%
%\enlargethispage{-1.2cm} 
%%
%% If the balancing should occur in the middle of the references, use
%% the following trigger:
%%
%\IEEEtriggeratref{4}
%%
%% which triggers a \newpage (i.e., new column) just before the given
%% reference number. Note that you need to adapt this if you modify
%% the paper.  The "triggered" command can be changed if desired:
%%
%\IEEEtriggercmd{\enlargethispage{-20cm}}
%%
%%%%%%

%%%%%%
%% References:
%% We recommend the usage of BibTeX:
%%
%\bibliographystyle{IEEEtran}
%\bibliography{definitions,bibliofile}
%%
%% where we here have assume the existence of the files
%% definitions.bib and bibliofile.bib.
%% BibTeX documentation can be obtained at:
%% http://www.ctan.org/tex-archive/biblio/bibtex/contrib/doc/
%%%%%%

%% Or you use manual references (pay attention to consistency and the
%% formatting style!):
 \bibliographystyle{IEEEtran}
   \bibliography{/Users/matteorosati/Documents/Ricerca/Appunti/library}

\end{document}